\documentstyle[titlepage,12pt]{article}
\newread\epsffilein    
\newif\ifepsffileok    
\newif\ifepsfbbfound   
\newif\ifepsfverbose   
\newdimen\epsfxsize    
\newdimen\epsfysize    
\newdimen\epsftsize    
\newdimen\epsfrsize    
\newdimen\epsftmp      
\newdimen\pspoints     
\def\epsfbox#1{\global\def\epsfllx{72}\global\def\epsflly{72}%
   \global\def\epsfurx{540}\global\def\epsfury{720}%
   \def\lbracket{[}\def\testit{#1}\ifx\testit\lbracket
   \let\next=\epsfgetlitbb\else\let\next=\epsfnormal\fi\next{#1}}%
\def\epsfgetlitbb#1#2 #3 #4 #5]#6{\epsfgrab #2 #3 #4 #5 .\\%
   \epsfsetgraph{#6}}%
\def\epsfnormal#1{\epsfgetbb{#1}\epsfsetgraph{#1}}%
\def\epsfgetbb#1{%
\openin\epsffilein=#1
\ifeof\epsffilein\errmessage{I couldn't open #1, will ignore it}\else
   {\epsffileoktrue \chardef\other=12
    \def\do##1{\catcode`##1=\other}\dospecials \catcode`\ =10
    \loop
       \read\epsffilein to \epsffileline
       \ifeof\epsffilein\epsffileokfalse\else
          \expandafter\epsfaux\epsffileline:. \\%
       \fi
   \ifepsffileok\repeat
   \ifepsfbbfound\else
    \ifepsfverbose\message{No bounding box comment in #1; using defaults}\fi\fi
   }\closein\epsffilein\fi}%
\def\epsfsetgraph#1{%
   \epsfrsize=\epsfury\pspoints
   \advance\epsfrsize by-\epsflly\pspoints
   \epsftsize=\epsfurx\pspoints
   \advance\epsftsize by-\epsfllx\pspoints
   \epsfxsize\epsfsize\epsftsize\epsfrsize
   \ifnum\epsfxsize=0 \ifnum\epsfysize=0
      \epsfxsize=\epsftsize \epsfysize=\epsfrsize
     \else\epsftmp=\epsftsize \divide\epsftmp\epsfrsize
       \epsfxsize=\epsfysize \multiply\epsfxsize\epsftmp
       \multiply\epsftmp\epsfrsize \advance\epsftsize-\epsftmp
       \epsftmp=\epsfysize
       \loop \advance\epsftsize\epsftsize \divide\epsftmp 2
       \ifnum\epsftmp>0
          \ifnum\epsftsize<\epsfrsize\else
             \advance\epsftsize-\epsfrsize \advance\epsfxsize\epsftmp \fi
       \repeat
     \fi
   \else\epsftmp=\epsfrsize \divide\epsftmp\epsftsize
     \epsfysize=\epsfxsize \multiply\epsfysize\epsftmp
     \multiply\epsftmp\epsftsize \advance\epsfrsize-\epsftmp
     \epsftmp=\epsfxsize
     \loop \advance\epsfrsize\epsfrsize \divide\epsftmp 2
     \ifnum\epsftmp>0
        \ifnum\epsfrsize<\epsftsize\else
           \advance\epsfrsize-\epsftsize \advance\epsfysize\epsftmp \fi
     \repeat
   \fi
   \ifepsfverbose\message{#1: width=\the\epsfxsize, height=\the\epsfysize}\fi
   \epsftmp=10\epsfxsize \divide\epsftmp\pspoints
   \vbox to\epsfysize{\vfil\hbox to\epsfxsize{%
      \includegraphics{#1}%
      \hfil}}%
\epsfxsize=0pt\epsfysize=0pt}%

{\catcode`\%=12 \global\let\epsfpercent=
\long\def\epsfaux#1#2:#3\\{\ifx#1\epsfpercent
   \def\testit{#2}\ifx\testit\epsfbblit
      \epsfgrab #3 . . . \\%
      \epsffileokfalse
      \global\epsfbbfoundtrue
   \fi\else\ifx#1\par\else\epsffileokfalse\fi\fi}%
\def\epsfgrab #1 #2 #3 #4 #5\\{%
   \global\def\epsfllx{#1}\ifx\epsfllx\empty
      \epsfgrab #2 #3 #4 #5 .\\\else
   \global\def\epsflly{#2}%
   \global\def\epsfurx{#3}\global\def\epsfury{#4}\fi}%
\def\epsfsize#1#2{\epsfxsize}
\textheight=23cm
\textwidth=15.0cm
\setlength{\topmargin}{-1cm}
\addtolength{\evensidemargin}{-0.5cm}
\addtolength{\oddsidemargin}{-0.5cm}
\newcommand{\be}{\begin{equation}}
\newcommand{\ee}{\end{equation}}
\newcommand{\bear}{\begin{eqnarray}}
\newcommand{\ear}{\end{eqnarray}}
\begin{document}
\begin{titlepage}
\begin{flushright}
PITHA 96/19\\
HD-THEP-96-20
\end{flushright}
\vspace{0.8cm}
\begin{center}
{\bf\LARGE Chiral-invariant  CP-violating Effective \\}
{\bf\LARGE Interactions in Z Decays to three Jets\footnote{
Research supported  by BMBF, contract No. 05 6HD 93P(6)}\\}
\vspace{2cm}
\centerline{ \bf W. Bernreuther$^a$, A.
Brandenburg$^{a,}$\footnote{Research supported by
Deutsche Forschungsgemeinschaft},
P. Haberl$^b$, and O. Nachtmann$^b$}
\vspace{1cm}
\centerline{$^a$Institut f. Theoretische Physik,
RWTH Aachen, D-52056 Aachen, Germany}
\centerline{$^b$Institut f. Theoretische Physik,
Universit\"at Heidelberg, D-69120 Heidelberg, Germany}
\vspace{3cm}
{\bf Abstract:}\\[2mm]
\parbox[t]{\textwidth}
{Tests of  CP violation by appropriate
momentum correlations  in $Z\to 3$ jets and in particular in
$Z\to b\bar bX$ probe CP-violating effective couplings -- that
manifest themselves as form factors -- 
which conserve the quark  chirality and quark flavour. By giving two examples 
we show that such couplings can be induced at one-loop order
in extensions of the Standard Model with CP violation beyond the
Kobayashi-Maskawa phase. In one of the models we compute  the  
chirality-conserving part of the CP-violating  $Zb{\bar b}$-gluon 
amplitude for massless $b$ quarks, determine the resulting 
effective dimension $d=6$ couplings in the local limit, and 
discuss the possible size of the effects. Finally we show that
in models with excited quarks the chiral-invariant CP-violating
effective interactions could be quite large if appropriate couplings 
are of a size characteristic of a strong interaction. }
\end{center}
\end{titlepage}
\newpage
\section{Introduction}
The large number of $Z$ boson events at LEP1 and at the SLAC 
linear collider have provided precision tests of the Standard Model 
(SM), and allow for searches of new physics effects. In \cite{1,2,3} 
it was proposed to use the reactions $Z\to n\ge 3$ jets and in 
particular $Z\to b{\bar b} X$ for tests of CP nonconservation 
beyond the one induced by the phase of the Kobayashi-Maskawa (KM) 
matrix \cite{4}. Related proposals were made in \cite{5,6,6a,6b}.
\par
CP-violating interactions in $Z\to b{\bar b} X$ would affect
at the parton level correlations among parton energies/momenta and 
parton spins. While the partonic momentum directions are reconstructed
from the jet directions of flight the spin-polarization of the $b$ 
quark (and of lighter quarks)  cannot, in general, be determined 
with reliable precision due to fragmentation. As far as CP tests 
in the above reactions are concerned this implies that useful 
observables are primarily those which originate from CP-odd partonic 
momentum correlations. With these correlations only chirality-conserving
effective couplings (which do not flip the quark helicity) can be  
probed with reasonable sensitivity \cite{1,2,3}. This situation is 
in contrast to $\tau^+\tau^-$ and $t\bar{t}$ production where 
the fermion polarizations can be traced in the decays. With regard to 
CP tests \cite{7} this circumstance allows to search in 
$Z\to \tau^+\tau^-$ for a CP-violating dipole form factor of the 
$\tau$ \cite{8,9} which is chirality-flipping.
\par
In the framework of a manifestly $SU(2)_L\times U(1)_Y$ gauge-invariant
effective Lagrangian approach it was shown in \cite{10} that 
 CP-violating interactions of dimension 
$d=6$ (after symmetry breaking) that are
chirality-conserving and flavour-diagonal  can exist with couplings which are 
a priori not related to the couplings of the $d=5$ dipole interactions.
In renormalizable theories these couplings can only be due to quantum
corrections. In this letter we show that there are extensions of the 
SM with CP violation beyond the KM phase where such ``couplings" -- 
that is to say, form factors -- are induced at one-loop order in the 
$Zb{\bar b}$-gluon amplitude even for vanishing $b$ quark mass. In one 
of the models discussed below we determine the resulting effective 
couplings in the ``local limit" where the masses of the particles 
in the loop are much larger than the external energies and discuss 
the possible strength of these couplings.
\par
Finally we make an  estimate for models with excited quarks and
show that  one can obtain rather large CP-violating and
chirality-conserving effective interactions relevant for 
$Z\to b\bar bX$. The price one has to pay is the introduction of
a new type of strong interactions for quarks.
\section{Parameterization of CP violation \newline in 
$Z\to b + {\bar b} + G$}
We consider the amplitude for the following
partonic reaction at the $Z$ resonance:
\begin{equation}
e^+ + e^- \to Z(p) \to {\bar b}(p_1) + b(p_2) + G(k)\;,
\label{reac}
\end{equation}
where $G$ denotes a gluon and the letters in brackets denote the
four-momenta of the particles. The amplitude (\ref{reac}) may be 
affected by CP-violating interactions. In the effective Lagrangian 
approach these interactions are parameterized by the couplings of local 
operators composed of the $Z$, $b$, $\bar b$, and gluon fields.
Including terms of operator dimension $d\le 6$ (counted after 
spontaneous breaking of the electroweak symmetry) the effective 
CP-violating Lagrangian relevant for (\ref{reac}) contains 
helicity-flipping $d=5$ electric, chromoelectric, and weak dipole 
moment operators and helicity-conserving $d=6$ operators. Here we 
are interested  only in the latter terms which read \cite{1,10}
\begin{equation}
{\cal L}_{\rm CP}^{d=6}(x) = {\bar b}(x)T^a\gamma^{\nu}
[h_{Vb}+h_{Ab}\gamma_5]b(x) Z^{\mu}(x) G^a_{\mu\nu}(x)\;,
\label{eff}
\end{equation}
where $T^a$ are the generators of $SU(3)_c$ in the fundamental 
representation, $G^a_{\mu\nu}$ denotes the gluonic field strength 
tensor, and $h_{Vb,Ab}$ are real coupling constants of mass dimension 
$-2$. The interactions (\ref{eff}) induce a CP-odd term in the 
amplitude of (\ref{reac}). Interference with the CP-even SM terms 
leads to CP-odd correlations among the momenta and momentum directions
of the $b$, $\bar b$, and gluon. The momenta and in particular the 
momentum directions are given to good accuracy by the corresponding 
jet variables. Detailed studies were made in \cite{1,2,3}. These 
CP-odd partonic and jet correlations, respectively, are not supressed 
by (powers of) $m_b/\sqrt s$. (Here $m_b$ denotes the $b$ quark mass 
and $\sqrt s$ is the c.m. energy.) It is convenient to work with
dimensionless coupling constants $\hat h_{Vb,Ab}$ which are defined by
\begin{equation}
h_{Vb,Ab}=\frac{eg_s}{\sin\theta_W\cos\theta_W m^2_Z}\hat h_{Vb,Ab}\;.
\label{hva}
\end{equation}
\noindent Here $e>0$ denotes the positron charge, $g_s$ is the QCD 
coupling constant, and $m_Z$ is the $Z$ boson mass.
\par
Alternatively one may parameterize CP-violating effects in (1) by 
performing a form factor decomposition of the amplitude (cf.\ 
\cite{6b} for an analysis of $Z\to b\bar{b}\gamma$ which also applies 
to (1)). For some remarks concerning the relation and the respective
advantages of the effective Lagrangian and the form factor approaches
we refer to \cite{120}. Restriction to on-shell $Z$ decays, massless 
$b$ quarks, and imposition of $SU(3)_c$ gauge invariance implies that 
the $Zb\bar{b}G$ vertex function contains  six pairs of
chirality-conserving form factors $(h^{(i)}_V + h^{(i)}_A
\gamma_5)$, $(i=1\ldots 6)$. One obtains for the CP-violating part 
of the ${\cal T}$-matrix element:
\begin{equation}
{i\cal T}_{CP} = \epsilon_Z^{\mu}\epsilon_G^{*a\nu}{\bar u}(p_2)
T^a\Lambda_{\mu\nu}v(p_1)\;,
\label{Tmat}
\end{equation}
where
\begin{equation}
\Lambda_{\mu\nu}=(/\hspace{-2mm}k g_{\mu\nu}-k_{\mu}\gamma_{\nu})
(h^{(1)}_V + h^{(1)}_A\gamma_5) + r_{\mu\nu}
\label{FF}
\end{equation}
and $\epsilon_Z^{\mu}$ ($\epsilon_G^{*a\nu}$) is the polarization
vector of the $Z$ boson (gluon) in ordinary (and colour) space.
The pair of form factors exhibited in (\ref{FF}) has mass dimension 
$-2$ and its accompanying Lorentz structure is identical to the 
structure induced by (\ref{eff}). The remainder $r_{\mu\nu}$ contains
other CP-odd form factor pairs which have mass dimension $-4$, i.e.,
their Lorentz structure contains two more powers of external momenta
\cite{6b}. In the local limit (cf.\ below) these form factors 
correspond to dimension $d\geq 8$ interactions, whereas 
$h^{(1)}_V$, $h^{(1)}_A$ become identical to the couplings
$h_{Vb}$, $h_{Ab}$, respectively, which are defined in (\ref{eff}).
\section{Extended Higgs models}
We now discuss extensions of the standard electroweak theory in which
chirality-conserving and CP-violating terms -- in particular terms with
a structure as described by (\ref{eff}), (\ref{FF}) --  are generated 
in the $Zb{\bar b}G$ amplitude  at one-loop order. The simplest 
possibility is the extension of the Standard Model by an additional 
Higgs doublet, which allows for CP violation in the neutral Higgs sector
\cite{11}. For 2-Higgs doublet models with natural flavour conservation 
at the tree level \cite{12} CP-violating neutral Higgs boson exchange at 
one-loop order leads to chirality-conserving structures in the above 
amplitude \cite{13}. However, these terms are proportional to $m^2_b$.
(For instance one factor of $m_b$ comes from the chirality-flipping 
Higgs boson coupling, the other one from the mass term necessary to 
flip the chirality back.) Thus the resulting effects at the $Z$ peak 
are uninterestingly small.
\par
Effects which do not vanish for $m_b\to 0$ are possible for instance
in models with $n > 2$ charged Higgs bosons and intrinsic CP 
violation in the charged Higgs sector (resulting from CP violation 
in the charged mixing matrix; i.e., the scalar potential). Moreover 
non-diagonal $ZH^+_iH^-_j$ couplings are required. They can appear 
if there are Higgs representations other than doublets. For 
definiteness we consider an extension of the standard 
$SU(2)_L\times U(1)_Y$ electroweak theory with three $SU(2)$ Higgs 
doublets and one singlet with $Y=1$. The physical particle spectrum 
of the model contains three charged Higgs bosons $H^{\pm}_{1,2,3}$ 
with an off-diagonal interaction with $Z$ bosons being of the
generic form
\begin{equation}
{\cal L}_{ZH_1H_2}= -i\frac{e}{\sin\theta_W\cos\theta_W}\kappa_{12} 
Z^\mu[H^+_1\partial_\mu H^-_2-(\partial_\mu H^+_1)H^-_2]+{\rm h.c.}
\label{ZHH}
\end{equation}
and similar terms with $H_1$, $H_3$ and $H_2$, $H_3$. Here 
$\kappa_{12}$ is a real parameter and $H^+_{1,2,3}$ denotes a physical 
charged Higgs boson in the mass basis. The interaction of one of these 
bosons with $t$ and $b$ quarks reads
\begin{equation}
{\cal L}_{H_itb}= -(2\sqrt 2G_F)^{1/2} (\alpha_i m_b {\bar t}_L b_R 
+ \beta_i m_t {\bar t}_R b_L) H^+_i + {\rm h.c.}
\label{Htb}
\end{equation}
Here $G_F$ denotes the Fermi constant and
$q_{R,L}=\frac{1}{2}(1\pm\gamma_5)q$. Due to CP-violating charged
Higgs boson mixing the phases of the complex numbers $\alpha_i$,
$\beta_i$ differ from the KM phase, and in general
${\rm Im}(\alpha_i\beta^*_i) \neq 0$. For generating non-zero form 
factors (\ref{FF}) in the limit of massless $b$ quarks
${\rm Im}(\beta_j\beta^*_k) \neq 0$ ($j\ne k$) is required. 
This is possible if $t_R$ has complex Yukawa couplings to more than 
two Higgs doublets. Such couplings  are not excluded for the third 
generation fermions. A detailed discussion of this model -- which we 
shall call model I for short -- will be given elsewhere.
\par
The interactions (\ref{ZHH}), (\ref{Htb}) generate the matrix 
element (\ref{Tmat}) at one-loop order if 
${\rm Im}(\beta_j\beta_k^*)\neq 0$ for $j\ne k$ and if the Higgs 
masses differ from each other. The diagrams of Fig.~1 depict the 
left-handed contribution to the vertex function which survives the 
limit $m_b\to 0$. The Yukawa interaction (\ref{Htb}) leads also to 
a right-handed vertex function which is, however, proportional to 
$m_b^2$ and therefore uninteresting. Putting the intermediate $Z$ 
boson on-shell\footnote{Since terms proportional to $p_\mu$ are 
irrelevant for the reaction (\ref{reac}) the result given below 
applies also to off-shell $Z$ bosons.} and the mass of the $b$ quark 
to zero, we find
that the left-handed vertex function represented 
by Figs.~1a, b contains all the CP-violating form factors
of mass dimension $-2$ and $-4$  mentioned above. 
 We are interested here only in the 
dimension $-2$ form factors for which we obtain the following
contribution from the $ZH_1H_2$ coupling in (6):
\begin{equation}
h^{(1)}_V = -h^{(1)}_A={\rm Im}(\beta_1\beta_2^*)\sqrt{2} G_F 
\frac{eg_s\kappa_{12}}{16\pi^2 \sin\theta_W\cos\theta_W} f_{12}\;.
\label{hvha}
\end{equation}
The amplitude $f_{12}$ is given as a function of the rescaled $b$ 
and $\bar{b}$ energies (in the $e^+e^-$ c.m.\ frame) $x=2E_b/\sqrt{s}$,
$\bar{x}=2E_{\bar{b}}/\sqrt{s}$ by
\begin{equation}
f_{12}(x,\bar{x})=\frac{m_t^2}{s}\int_0^1dx_1\int_0^{1-x_1}dx_2
\left[I_a(x,\bar{x},x_1,x_2)+I_b(x,\bar{x},x_1,x_2)\right]\;.
\label{intf}
\end{equation}
With
\begin{equation}
\rho(m_{1},m_{2})\equiv \frac{x_1m_{2}^2+x_2m_{1}^2
+(1-x_1-x_2)m_t^2}{s}\;,
\end{equation}
where $m_{1,2}$ denote the charged Higgs masses, we have
\begin{eqnarray}
I_a &=& \frac{2x_1x_2(1-x-\bar{x})-x_2(1-x)-x_1(1-\bar{x})
+2\rho(m_{1},m_{2})}{[x_2(1-x)-x_1(1-\bar{x})]^2}
\nonumber \\[2mm]
&& \times \; \ln\left[ \frac{\rho(m_{1},m_{2})-x_2(1-x_2)(1-x)
-x_1x_2x}{\rho(m_{1},m_{2})-x_1(1-x_1)(1-\bar{x})-x_1x_2\bar{x}}
\right] \\[2mm] &-& (m_{1}\leftrightarrow m_{2}) \nonumber
\end{eqnarray}
and
\begin{eqnarray}
I_b &=& \frac{\ln\left[\rho(m_{1},m_{2})-x_1x_2-x_1(1-x_1-x_2)
(1-\bar{x})\right]}{1-\bar{x}}\nonumber \\[2mm]
&-& \frac{\ln\left[\rho(m_{1},m_{2})-x_1x_2+(1-x_2)(1-x_1-x_2)
(1-x)\right]}{1-x} \label{Ib}\\[2mm]
&-& (m_{1}\leftrightarrow m_{2})\;. \nonumber
\end{eqnarray}
In (\ref{intf}) $I_a$ ($I_b$) is the contribution from the one 
particle irreducible (reducible) diagram(s) in Fig.~1. Correspondingly,
$I_b$ could in principle have poles at $x=1$ and $\bar x=1$. However,
in (\ref{Ib}) these pole terms cancel when the contribution 
$(m_{1}\leftrightarrow m_{2})$ is subtracted. Thus, $h_V^{(1)}$,
$h_A^{(1)}$ in (\ref{hvha}) are regular for $x\to 1$, $\bar x\to 1$
and in the local limit discussed below they get equal contributions 
from the diagrams (a) and (b) of Fig.~1.
\par
The form factor $h^{(1)}_V$ does not show any pronounced variation 
as a function of $x$, $\bar{x}$ in the kinematically allowed range.
In order to make contact with the effective Lagrangian approach we  
assume that the particle masses in the loops of Fig 1a, b are much 
larger than the external momenta involved. This local limit can 
formally be realized by assuming $s\ll m_t^2$, $m_1^2$, $m_2^2$ in 
(\ref{intf}). With $r_{1,2}\equiv m_{1,2}^2/m_t^2$ we obtain then
\newpage
\begin{eqnarray}
f_{12}(x,\bar{x}) \to {\hat f}(r_1,r_2) \;\;=\;\; 
\frac{2}{3(r_1-r_2)} \!\!\!\!&\Bigg\{&\!\!\!\!
- \;\;\frac{2[(r_1-r_2)^2+r_1r_2(1-r_1)(1-r_2)]}{(1-r_1)^2(1-r_2)^2}
\nonumber \\ 
- \;\;\frac{r_1^2\ln(r_1)(r_1-3r_2+r_1r_2+r_1^2)}{(1-r_1)^3(r_1-r_2)} 
\!&+&\! \frac{r_2^2\ln(r_2)(r_2-3r_1+r_1r_2+r_2^2)}
{(1-r_2)^3(r_1-r_2)}  \Bigg\}\;.  \label{local} 
\end{eqnarray}
The function ${\hat f}(r_1,r_2)$ is an antisymmetric and regular
function of the mass ratios $r_{1,2}$. The maximal value of its 
modulus is $|{\hat f}(1,0)| = |{\hat f}(0,1)| = 1/9$. From Fig.~2 
we see that $|{\hat f}(r_1,r_2)| \simeq  0.07$ if one Higgs mass 
is around 90 GeV and the other one is around 300 GeV or larger.
Furthermore a numerical study shows that replacing, for a given set of 
particle masses, the form factor function $f_{12}(x,\bar x)$ by its 
local limit $\hat f(r_1,r_2)$ is a very good approximation 
in the whole kinematic range of $x$, $\bar x$.
\par
We can  now estimate the strength of the effective interaction 
(2). It is characterized by the dimensionless couplings
(3) for which we get from (\ref{hvha})--(\ref{local})
after summing the contributions of all pairs $H_j$, $H_k$ 
($j\ne k$)
\begin{equation}
{\hat h}_{Vb}=-{\hat h}_{Ab}=\sum_{1\le j<k\le 3}
\sqrt{2} G_F m_Z^2\frac{\kappa_{jk} {\rm Im}(\beta_j\beta_k^*)}
{16 \pi^2} {\hat f}(r_j,r_k)\;.
\label{hhb}
\end{equation}
How large could $\hat h_{Vb,Ab}$ possibly be? The parameters 
$\kappa_{jk}$ will in general not exceed values of order one. If the 
couplings of the charged Higgs bosons to the right-handed top quark 
are substantially enhanced, $|\beta_j| = {\cal O}(10)$, then 
$\hat h_{Vb,Ab}$ can reach the per cent level.
\par
Models of another class in which a non-zero matrix element 
(4) can be induced at one-loop order are $SU(2)_L\times U(1)_Y$
gauge theories with exotic Higgs representations. It is known 
[17,18] that models with Higgs boson multiplets other than 
doublets or singlets  can have tree level $H^{\pm}W^{\mp}Z$ 
couplings ($H^+$ denotes again a physical charged Higgs boson
in the mass basis) which may be parameterized as follows [18]:
\begin{equation}
{\cal L}_{HWZ}= -\frac{e}{\sin\theta_W \cos\theta_W} m_Z \xi 
W^+_\mu Z^\mu H^- + {\rm h.c.}
\label{HWZ}
\end{equation}
Here $\xi$ is a real parameter which depends on the vacuum expectation 
values and on the quantum numbers of the scalar fields. It should be 
noted that the experimental constraint of the $\rho$ parameter being 
close to 1 does not imply that $\xi$ must be very small. In these 
models there is usually more than one  singly charged physical Higgs 
state. Depending on the scalar potential there can be CP-violating
mixing of these states. Then  the couplings of the charged Higgs bosons
to quarks can have a CP-violating phase being different from the KM 
phase. The interaction of one of these bosons $H^+$ is of the generic 
form (7) with complex couplings $\alpha,\beta$.
\par
If ${\rm Im}(\beta V_{tb}^*)\neq 0$, where $V_{tb}$ is the KM mixing
matrix element, these interactions -- which we call model II --
generate the left-handed  amplitude depicted in Fig.~3a, b with 
CP-violating form 
factors which remain non-zero for $m_b = 0$. 
In the local limit $s/m^2_i\to 0$ $(i=H,W,t)$ we obtain for the 
dimensionless couplings (3)
\begin{equation}
{\hat h}_{Vb}=-{\hat h}_{Ab}=\sqrt{2} G_F m_Z^2
\frac{\xi {\rm Im}(\beta V_{tb}^*)}{16 \pi^2} \tilde f\;,
\label{hhhb}
\end{equation}
where $\tilde f$ is the local limit of the form factor function 
associated with $h^{(1)}_V$. It is of the same order of magnitude 
as $\hat f$ in (\ref{local}).
We may expect the absolute value of
$\hat h_{Vb}$ evaluated in model II to be smaller than the one of 
 $\hat h_{Vb}$ in model I because for 
enhanced  Yukawa coupling $\beta$ the coupling 
(\ref{hhhb}) grows only linearly with $\beta$.
\par
In this section we have discussed in detail only the form factors 
$h^{(1)}_V$, $h^{(1)}_A$ and their local limits. When comparing the 
predictions of these models with experimental data, the other
CP-violating form factors of mass dimension $-4$
should also be taken into 
account. In this note our intention was to show that in the context
of spontaneously broken $SU(2)_L\times U(1)_Y$ gauge theories with 
CP violation beyond the KM phase there can be CP-violating 
contributions to $Z\to b+\bar b + G$ at one-loop approximation which 
conserve the quark chirality and quark flavour.
 In the (possibly hypothetical) limit 
where the particle masses in the loop are much larger than the 
external momenta these contributions lead in particular to dimension 
$d = 6$ chiral-invariant effective quark gauge-boson interactions. 
The corresponding couplings are not suppressed by small quark masses. 
\section{Models with excited quarks}
Let us assume that there exist excited quarks
(cf.~e.g.\ \cite{180}). This would be natural
in a scenario where quarks have substructure and participate in a
new type of strong interaction. In particular, we assume that $b$
quarks have excited partners $b'$, which could have spin 
$\frac{1}{2}$ or $\frac{3}{2}$. For simplicity we consider a $b'$ 
of spin $\frac{1}{2}$ and mass $m_{b'}$. Due to
colour gauge invariance we expect the $b'b$ Gluon couplings to be
chirality-flipping dipole couplings. Because weak $SU(2)$ gauge
invariance is broken at the scale of LEP energies
chirality-conserving $Zb'b$ couplings are a priori
possible. For the sake of demonstrating that
couplings of the form given in (2) can be generated
we consider the following effective interactions of $b'$ to $b$ quarks,
$Z$ bosons and gluons:
\begin{equation}
{\cal L}'=-\;\frac{e}{2\sin\theta_W \cos\theta_W}Z_\mu\bar b'
\gamma^\mu(g'_V-g'_A\gamma_5)b - i \frac{g_s}{2m_{b'}}\hat{d}_c
\bar b'\sigma^{\mu\nu}\gamma_5T^abG^a_{\mu\nu}+{\rm h.c.}
\label{exq}
\end{equation}
Here $g'_V$, $g'_A$ and $\hat{d}_c$ are complex parameters, which 
can be expected 
to be of order one if the underlying dynamics is strongly interacting.
In addition to  $\hat{d}_c$, the 
chromoelectric dipole transition form factor $b\to b'$, there will 
be also a chromomagnetic transition form factor $\hat{d}_m$ which we omit
for  brevity.

It is a simple matter to calculate $\hat h_{Vb,Ab}$ in this type 
of model from the diagrams of Fig.~4.  For $m_{b'}\gg m_Z$ we get
\begin{eqnarray}
\hat h_{Vb}&=&\hphantom{-}\;\frac{m_Z^2}{m^2_{b'}}
{\rm Re}(\hat{d}_c g_A^{'*})\;,\nonumber \\
\hat h_{Ab}&=&-\;\frac{m_Z^2}{m^2_{b'}}{\rm Re}(\hat{d}_c g_V^{'*})\;.
\end{eqnarray}
If $m_{b'}$ is not too far away from $m_Z$ the couplings
 $\hat h_{Vb,Ab}$ get somewhat enhanced due to
the propagator of the virtual $b'$ quark in Fig.~4. 
In addition  the contribution of the chromomagnetic 
transition coupling $\hat{d}_m$ to  $\hat h_{Vb}$, $\hat h_{Ab}$ 
is proportional to
${\rm Im}
(\hat{d}_m g_V^{'*})$ and ${\rm Im}(\hat{d}_m g_A^{'*})$, respectively.
Thus we conclude 
that in this model $\hat h_{Vb,Ab}={\cal O}(1)$ may be possible 
if the mass of the $b'$ quark is not too far away
from the $Z$ mass. In \cite{181} the lower limit $m_{q'} >$ 540 GeV 
on  the masses of excited quarks
was published, but this applies to excited $u$ and $d$ quarks only
and does not exclude a lighter $b'$ quark.
\par
In \cite{3} it was pointed out that the CP-odd couplings $(\ref{eff})$,
which enhance the 
$Z\to b\bar bG$ decay rate, would explain the  discrepancy of the 
experimental value of 
$R_b=\Gamma(Z\to b\bar bX)/\Gamma(Z\to {\rm hadrons})$ with the 
corresponding theoretical SM value  (cf.\ \cite{100}) 
if  they were of order one. Our present study indicates that couplings 
of this order of magnitude might be 
generated if there  are sufficiently light 
excited $b$ quarks. Of course, $R_b$ receives not only incoherent
contributions  from $\hat h_{Vb,Ab}$ but also coherent ones from the 
interference
of CP-invariant contributions induced by (\ref{exq}) with the SM contributions 
to the amplitude of
$Z\to b{\bar b} G$. 
These coherent contributions to $R_b$ 
could be more important than the incoherent ones and 
they  may also be positive. 

We add a remark concerning $b'$ production at the Tevatron collider.
Depending on the $b'$ mass, the couplings (\ref{exq}) may lead 
to anomalous ${\bar b}b'$ production, that is, anomalous ${\bar b}bG$ 
production with subsequent high $p_T$ dileptons 
from semileptonic $b$ and $\bar b$ decay.

\section{Conclusions}
We have discussed in this letter some ways to generate chiral-invariant
CP-violating couplings relevant for the decay $Z\to b\bar bG$. We
found that models with extra charged Higgs particles can induce such
couplings which survive the $m_b\to 0$ limit. The resulting 
dimensionless coupling constants $\hat h_{Vb,Ab}$ were estimated to 
be at most at the per cent level for realistic Higgs masses and Yukawa
couplings. Larger couplings $\hat h_{Vb,Ab}$
can be obtained in models with an excited $b$ quark, $b'$, with a mass 
not too far away from the $Z$ boson mass.
\par
We think that further experimental
investigations of $Z\to b\bar bG$ (and also $Z\to b\bar b\gamma$)
should be a very worthwhile undertaking. First studies in this 
direction 
have already been presented for $Z\to b\bar bG$ in \cite{101,102,103}.
\subsection*{Acknowledgements}
The authors are grateful to S.~Dhamotharan, J.~von Krogh, 
P.~Overmann, M.~Steiert, H.~Stenzel, M.~Wunsch, and P.~Zerwas 
for many useful discussions.
\vfil\eject
\section*{Figure Captions}
\begin{description}
\item{Fig.~1:} CP-violating  contributions to the $Zb{\bar b}G$ amplitude 
in model I for massless $b$ quarks. Permuted diagrams and diagrams where 
the gluon is emitted from the $\bar b$ quark are not drawn.
\item{Fig.~2:} Dependence of $\hat f_{12}$ defined in eq. (\ref{local}) as 
a function of the charged Higgs masses $m_1$ and $m_2$. The top mass is 
chosen to be $m_t$ = 175 GeV.
\item{Fig.~3:} CP-violating  contributions to the $Zb{\bar b}G$ amplitude
in model II for massless $b$ quarks.
\item{Fig.~4:} Contribution to $Zb{\bar b}G$
from an excited quark $b'$. The permuted diagram is not shown.
\end{description}

\vfil\eject
\setlength{\unitlength}{1cm}
\begin{picture}(15,10)
\hskip-0.5cm 
\epsfysize=8cm
\epsfbox{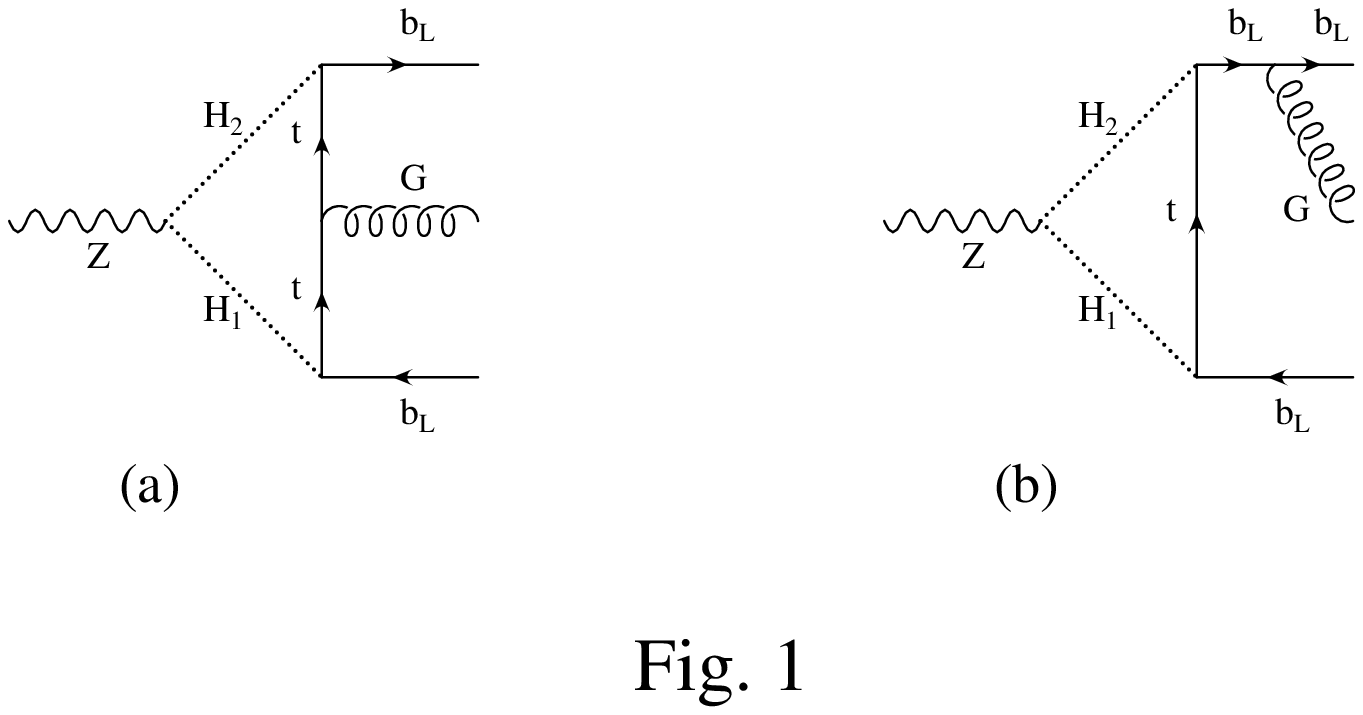}
\end{picture}
\ 
\vskip-2.75cm
\ 
\ 
\vfil\eject
\ 
\vskip 7cm
\begin{picture}(15,10)
\hskip -1cm 
\epsfysize=22cm
\epsfbox{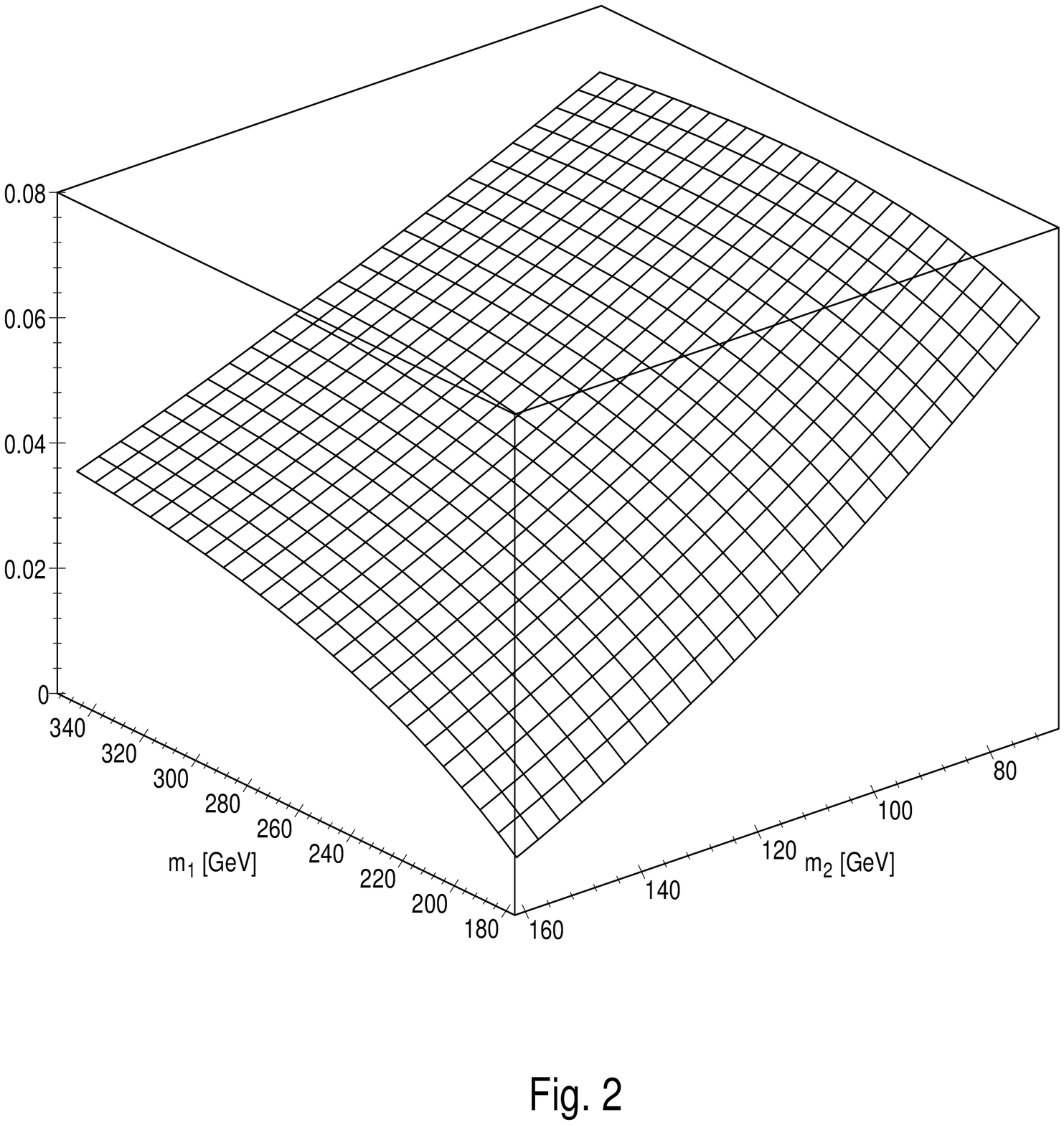}
\end{picture}
 
\ 
\
\vskip 0.75cm
\vfil\eject 
\begin{picture}(15,10)
\hskip -0.5cm 
\epsfysize=8cm
\epsfbox{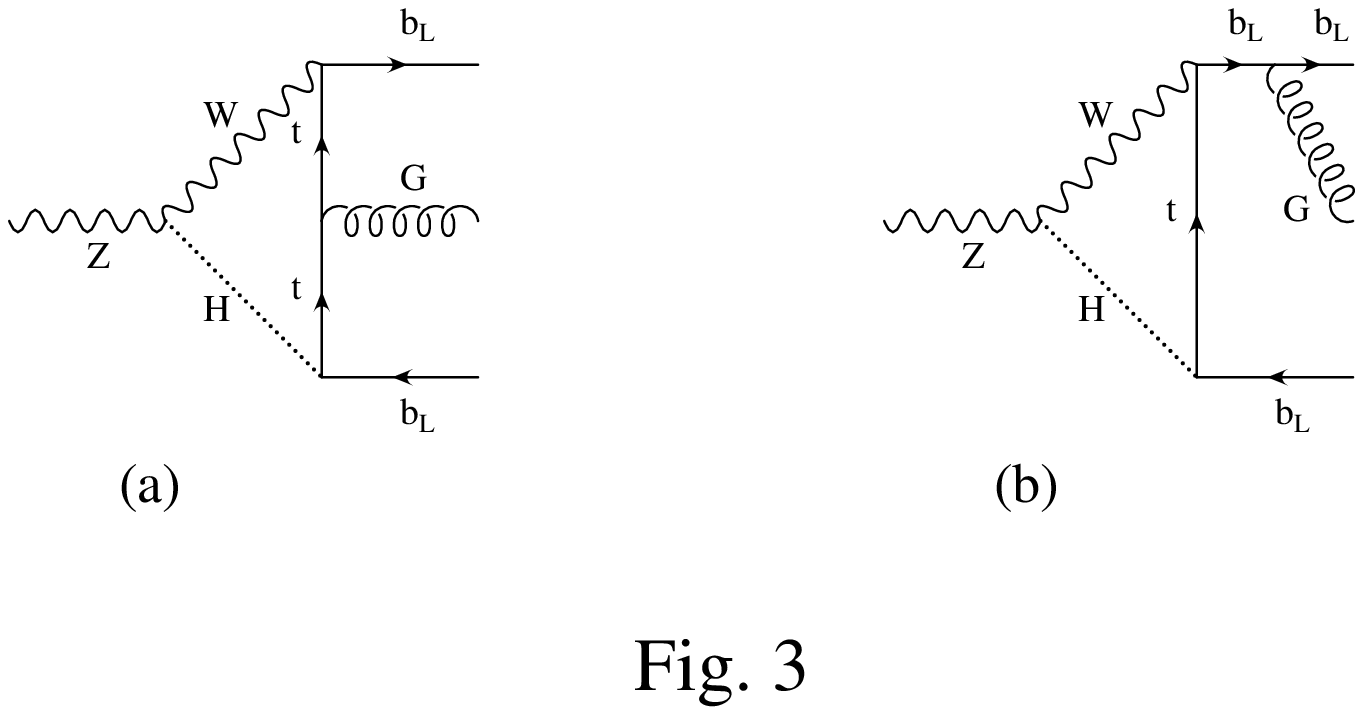}
\end{picture}
\ 
\vskip -2.75cm
\ 
\ 
\vskip 3.5cm

\begin{picture}(15,10)
\hskip 5.25cm 
\epsfysize=8cm
\epsfbox{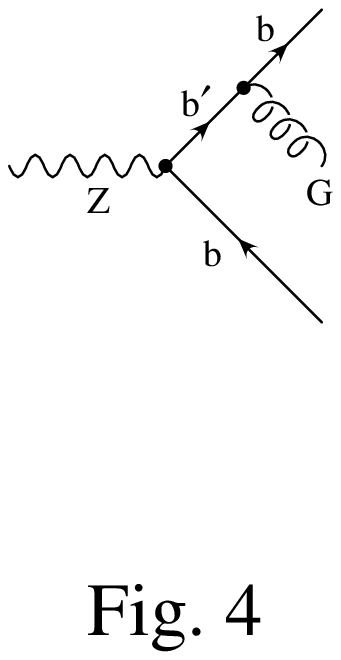}
\end{picture}
\ 
\vskip -2.75cm
\  

\end{document}